# Conditional Phase Shift for Quantum CCNOT Operation

[1]Miroshnichenko G.P., [2]Trifanov A.I.

Saint-Petersburg State University of Information Technologies, Mechanics and Optics
197101, Kronverksky 49, Saint-Petersburg, Russia

[1]gpmirosh@gmail.com,
[2]alextrifanov@gmail.com.

## Abstract

We suggest the improvement of description methods for quantum phase gate implementation in the cavity of QED configuration. Qubits are encoded into two lowest Fock state. Three qubit phase transformation is resulted from the interaction between Rydberg atom and three modes of cavity electromagnetic field. Evolution of conditional field states appearing after atom measurement is described by Kraus transformers. One of these operators is very convenient for phase gate implementation. We show that it corresponds to conditional evolution without quantum jumps. Also we describe cavity based generating EPR pair from certain initially disentangled state.

## 1. Introduction

In recent years, quantum information technologies [1, 2, 3, 4] are desired goal for engineers and scientists who work on quantum physics applications. Exponential acceleration in quantum algorithms [5], best security in quantum information transmission [6] and dense coding are based on quantum mechanics concepts [7]. Subject to physical system photonic and NMR technologies, trapped atoms and ions, superconductivity and quantum dots are proposed for quantum bit (qubit) implementation [1]. Processing of quantum information takes place in quantum computer which hardware must satisfy the requirements of scalability, controllability and correctability [8]. In addition the universal sets of quantum gates are needed. This means that such devises may form the logical basis. Two-qubit Fredkin CNOT (Controlled NOT) gate together with single qubit Hadamard operation satisfy universality requirement, as well as three-qubit Toffoli gate CCNOT (Controlled-Controlled NOT).

In optical quantum technologies [9, 10] photonic degrees of freedom are used for encoding the qubit state. Due to slow interaction between photons in free space they consider as a best carriers of information in quantum transmitting channels. On the other hand this advantage turns out to be an obstacle in implementation of many-qubit gates. There are two way under consideration here. In linear optic quantum computing (LOQC) [11] photon interference is used. Information processing obtained in optical schemes which consist of linear elements only. In quantum gates based on nonlinear interactions electromagnetic induced transparency (EIT) [12], giant Kerr effect [13, 14] and interaction in cavity with QED configuration [15] are usually used. Optical implementation of universal many-qubit gates CNOT and CCNOT based on Conditional Phase Shift (CPS) operation [16, 17, 18]. This quantum transformation multiplies every component of many-qubit state on phase coefficient dependent on the states of separate qubits. The devices which perform such operation are called Quantum Phase Gates (QPG). Universality of n-qubit quantum gate based on CPS requires the following condition to be satisfied:

$$Q_\pi^{(n)} |\alpha_1\rangle_1 |\alpha_2\rangle_2 \ldots |\alpha_n\rangle_n = \exp\left(-i\pi \delta_{\alpha_1}^1 \delta_{\alpha_2}^1 \ldots \delta_{\alpha_n}^1\right) |\alpha_1\rangle_1 |\alpha_2\rangle_2 \ldots |\alpha_n\rangle_n. \tag{1}$$

Here $Q_\pi^{(n)}$ is an operator corresponded to CPS, $\alpha_j \in \{0,1\}$, $j = 1,\ldots,n$ are quantum states of separate qubits in computational basis and $\delta_j^k$ is Kronecker delta. Then in the case of $n = 2$ and 3 the CNOT and CCNOT gates may be obtained after following transformations:

$$CNOT = (I \otimes H) \cdot Q_\pi^{(2)} \cdot (I \otimes H), \qquad (2)$$

$$CCNOT = (I \otimes I \otimes H) \cdot Q_\pi^{(3)} \cdot (I \otimes I \otimes H), \qquad (3)$$

where $I$ is identity operator and $H$ is single-qubit Hadamard operation.

In the QED-cavity the quantum modes of electromagnetic field interact with the circular Rydberg levels (principle quantum numbers $n = 50$ or $n = 51$) of atom passing through the cavity. The typical radiative time of such levels (30 ms) and photon storage time (1 ms) a several time greater than atom-field interaction in the cavity (10 μs). It allows obtaining the longlived entanglement between atom and field state [19]. This provides a powerful instrument for quantum logical gate implementation. The Hadamard operation [20], two- [20, 21] and three-qubit [22] phase gates implementation based on high Q-cavity were proposed. Here the qubits were represented by single cavity modes restricted in the space spanned by the two lowest Fock state. In [23] the circular Rydberg levels of atom were suggest for encoding qubit state. The logical gates are proposed to be implemented in the system which consists of several high-Q cavities [15]. In [24] the unconventional geometric phase gates were proposed.

In present work we suggest the improvement of methods which used for description of quantum logical gates based on conditional field state. This approach is used for implementation of $Q_\pi^{(3)}$ transformation. It is resulted from the interaction between six-levels Rydberg atom with three modes of electromagnetic field exited in the cavity of QED-configuration. Leaving the cavity atom goes to the detector (ionizing camera) where its state is measured. Conditional evolution of the field obtained after atomic measurement may be described using conditional field operator (Kraus transformer [25]). We used the evolution operator expansion to obtain the operator-valued system of differential equations for Kraus operators. Then all modes of electromagnetic field are in resonance with corresponding atomic transitions this system allows the exact analytical solutions for all transformers. Among the set of Kraus operators obtained for defined initial conditions for atom and field there is one which corresponds to conditional field state evolution when after interaction atom is detected in its initial state. Such operator is diagonal in the basis of Fock states and ideal for CPS implementation. Other Kraus operators describe the quantum jumps, when the photon number doesn't conserve. Using one of these "jump" operators we will denote one extra facility of the system we investigated. Namely, we will show that it is possible to create entangled EPR state from some initially disentangled state. In spite of the fact that the probability of such transformation is small (about 0.2) this operation may be useful for quantum communications.

The structure of this paper is organized as follows: in section 2 we describe the atom-filed system which will be used for $Q_\pi^{(3)}$ implementation; section 3 introduces the operators of conditional field evolution and solution of Schrodinger equation for them; section 4 is devoted to $Q_\pi^{(3)}$ implementation, here we give the estimations for time interaction, operation probability and fidelity; in section 5 we suggest the way of generation EPR state and section 6 conclude the article.

## 2. A model

The system we a going to consider consists of the six atomic levels which interact with three quantum and two classical modes of cavity electromagnetic field (see Fig. 1). Atomic population is assumed to be initially in the ground state $|1\rangle_A$. From this ground state, it could be excited by the first quantum field with coupling constant $g_a$ to the state $|2\rangle_A$. Two classical fields with Rabi frequencies $\Omega_1$ and $\Omega_2$ couple the atomic transitions $|2\rangle_A \leftrightarrow |3\rangle_A$ and $|2\rangle_A \leftrightarrow |5\rangle_A$ correspondingly. Both transitions $|3\rangle_A \leftrightarrow |4\rangle_A$ and $|5\rangle_A \leftrightarrow |6\rangle_A$ are driven by the second (b) and third (c) quantum modes with coupling constants $g_b$ and $g_c$. The Hamiltonian operator $H$ of the whole system is:

$$H = H_A + H_F + V. \tag{4}$$

Here $H_A = \sum_{k=1}^{6} E_k \sigma_{kk}$ and $H_F = \sum_{m \in \{a,b,c\}} \hbar \omega_m a_m^+ a_m$ are Hamiltonian operators of pure atomic and field subsystems. Atomic projectors $\sigma_{kk} = |k\rangle_A \langle k|$, $k \in \{1,2,\ldots,6\}$ act onto the subspace generated by eigenvalues $E_k$, $a_m$ and $a_m^+$, $m \in \{a,b,c\}$ are annihilation and creation operators of m-th mode. Interaction operator $V = V_c + V_q$ consists of two parts which correspond to classical and quantum fields ( $h.c.$ is Hermitian conjugation, here and further $\hbar = 1$ ):

$$V_c = -\Omega_1 \sigma_{23} \exp(i\omega_1 t) - \Omega_2 \sigma_{25} \exp(i\omega_2 t) + h.c., \tag{5}$$

$$V_q = -g_a\left(\sigma_{21} a_a + \sigma_{12} a_a^+\right) - g_b\left(\sigma_{43} a_b + \sigma_{34} a_b^+\right) - g_c\left(\sigma_{65} a_c + \sigma_{56} a_c^+\right). \tag{6}$$

Then in dipole and rotating wave approximation Hamiltonian operator may be written as follows:

$$H = \sum_{k=1}^{6} \Delta_k \sigma_{kk} - \left(g_a \sigma_{21} a_a + g_b \sigma_{43} a_b + g_c \sigma_{65} a_c + h.c.\right) - \left[\Omega_1 \sigma_{23} + \Omega_2 \sigma_{25} + h.c.\right]. \tag{7}$$

Here $\Delta_i$ $i \in \{1,2,\ldots,5\}$ are photon detunings. They may be expressed using single photon detunings $\varepsilon_i$ in the following way:

$$\Delta_1 = \varepsilon_1, \ \Delta_2 = \varepsilon_1 - \varepsilon_2, \ \Delta_3 = \varepsilon_1 - \varepsilon_2 + \varepsilon_3,$$
$$\Delta_4 = \varepsilon_1 - \varepsilon_4, \ \Delta_5 = \varepsilon_1 - \varepsilon_4 + \varepsilon_5. \tag{8}$$

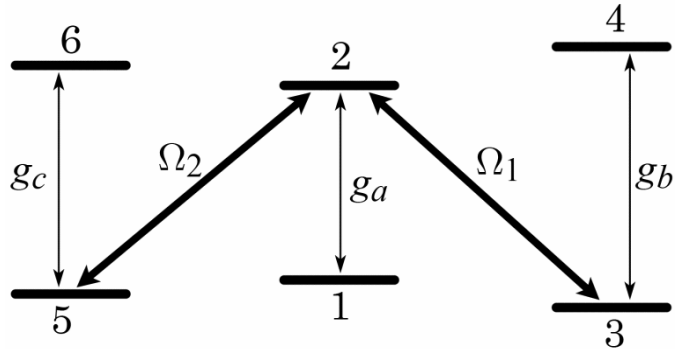

**Fig. 1:** Six levels energy structure of Rydberg atom passing through the QED cavity. $\Omega_{1,2}$ are Rabi frequencies of classical fields and $g_{a,b,c}$ are coupling constants of quantum cavity fields.

## 3. Conditional field evolution

Here we will consider the time evolution of the atom-field system using operator $U(t)$. It satisfies the following differential equation:

$$i\frac{\partial}{\partial t}U(t) = HU(t), U(0) = I. \qquad (9)$$

To this end let us expand $U(t)$ in the following way:

$$U(t) = \sum_{k,j=1}^{6} |k\rangle_A \langle j| K_{kj}(t), \qquad (10)$$

where $|k\rangle_A \langle j|$ are atomic projectors and $K_{kj}(t)$ are Kraus field operators. Each of them corresponds to conditional field evolution when atom initially in state $|j\rangle_A$ after interaction is measured in state $|k\rangle_A$, $j,k \in \{1,\dots,6\}$. In our system $|k\rangle_A = |1\rangle_A$ and we can write:

$$U(t) = \sum_{j=1}^{6} |j\rangle_A \langle 1| K_{j1}(t). \qquad (11)$$

Substituting (11) in (9) we obtain the following operator-valued system of differential equations for $K_j(t) \equiv K_{j1}(t)$:

$$\begin{aligned}
i\dot{K}_1(t) &= \Delta_1 K_1(t) + g_a^* a_a^+ K_2(t), \\
i\dot{K}_2(t) &= g_a a_a K_1(t) + \Delta_2 K_2(t) + \Omega_1^* K_3(t) + \Omega_2^* K_5(t), \\
i\dot{K}_3(t) &= \Omega_1 K_2(t) + \Delta_3 K_3(t) + g_b^* a_b^+ K_4(t), \\
i\dot{K}_4(t) &= g_b a_b K_3(t) + \Delta_4 K_4(t), \\
i\dot{K}_5(t) &= \Omega_2 K_2(t) + \Delta_5 K_5(t) + g_c^* a_c^+ K_6(t), \\
i\dot{K}_6(t) &= g_c a_c K_5(t) + \Delta_6 K_6(t).
\end{aligned} \qquad (12)$$

Under the assumption $\Delta_i = 0$, $i \in \{1,2,\dots,5\}$ it is simple to obtain the six-order homogeneous differential equation for Kraus operator $K_1(t)$:

$$K_1^{(6)}(t) - pK_1^{(4)}(t) + qK_1^{(2)}(t) - rK_1(t) = 0. \qquad (13)$$

Here upper index in the brackets denote the corresponding derivation of $K_1(t)$ and $p$, $q$, $r$ are time independent operator-value coefficients:

$$\begin{aligned}
p &= |g_a|^2 a_a^+ a_a + |g_b|^2 a_b^+ a_b + |g_c|^2 a_c^+ a_c + |\Omega_1|^2 + |\Omega_2|^2, \\
q &= |g_b|^2 \left(|\Omega_2|^2 + |g_a|^2 a_a^+ a_a\right) a_b^+ a_b + |g_c|^2 \left(|\Omega_1|^2 + |g_a|^2 a_a^+ a_a\right) a_c^+ a_c + |g_b|^2 |g_c|^2 a_b^+ a_b a_c^+ a_c, \\
r &= |g_a|^2 |g_b|^2 |g_c|^2 a_a^+ a_a a_b^+ a_b a_c^+ a_c.
\end{aligned} \qquad (14)$$

It should be noted that operators $p$, $q$ and $r$ are diagonal in the basis of Fock states. According to initial conditions $K_1(0)$ is also diagonal in this basis. From here we can conclude that Kraus operator $K_1(t)$ is diagonal permanently. It corresponds to conditional evolution of the cavity field and we will use it for $Q_\pi^{(3)}$ (1) implementation. Other operators $K_{k\neq 1}(t)$ don't conserve the photon number in cavity modes and, consequently, represent quantum jumps. Due to $K_1(t)$ is diagonal the solution of (13) may be found in the standard form:

$$K_1(t) = \sum_{j=1}^{6} C_j \exp(-i\Lambda_j t). \tag{15}$$

Here $C_j$ is the time independent operators which commutes with operator-valued root $\Lambda_j$ of characteristic equation obtained from (13):

$$\Lambda^6 - p\Lambda^4 + q\Lambda^2 - r = 0. \tag{16}$$

To express $C_j$ in terms of $\Lambda_j$ we use the following initial conditions for Kraus field operators:

$$K_1(0) = I, \ K_{k\neq 1}(0) = 0. \tag{17}$$

The system obtained from here may be separated by two parts:
a) inhomogeneous part for $B_1 = C_1 + C_2$, $B_2 = C_3 + C_4$ and $B_3 = C_5 + C_6$:

$$\begin{aligned} B_1 + B_2 + B_3 &= 1, \\ B_1\Lambda_1^2 + B_2\Lambda_2^2 + B_3\Lambda_3^2 &= |g_a|^2 a_a^+ a_a, \\ B_1\Lambda_1^4 + B_2\Lambda_2^4 + B_3\Lambda_3^4 &= |g_a|^2 \left( |g_a|^2 a_a^+ a_a + |\Omega_1|^2 + |\Omega_2|^2 \right) a_a^+ a_a, \end{aligned} \tag{18}$$

and
b) homogeneous part for $A_1 = C_1 - C_2$, $A_2 = C_3 - C_4$ and $A_3 = C_5 - C_6$:

$$\begin{aligned} A_1\Lambda_1 + A_2\Lambda_2 + A_3\Lambda_3 &= 0, \\ A_1\Lambda_1^3 + A_2\Lambda_2^3 + A_3\Lambda_3^3 &= 0, \\ A_1\Lambda_1^5 + A_2\Lambda_2^5 + A_3\Lambda_3^5 &= 0. \end{aligned} \tag{19}$$

Here we assume that all operator-valued roots of (16) are different. This gives the trivial solution of system (19). Then we can write:

$$K_1(t) = \sum_{j=1}^{3} B_j \cos(\Lambda_j t), \tag{20}$$

where $B_l$, $l \in \{1,2,3\}$ are the solutions of the inhomogeneous system. The elements of $K_1(t)$ in restricted Fock basis may be obtained explicitly (for simplification we set $|g_a|^2 = |g_b|^2 = |g_c|^2 = |g|^2$). Let us denote $K_1(t)_{s,s} = \langle s|K_1(t)|s\rangle_F$, where $|s\rangle_F = |\alpha_a \alpha_b \alpha_c\rangle_F$, $\alpha_m \in \{0,1\}$, $m \in \{a,b,c\}$, $s \in \{0,\ldots,7\}$ is the decimal notation of cavity field state in computational basis. It gives the following expressions for matrix elements:

$$K_1(t)_{00} = K_1(t)_{11} = K_1(t)_{22} = K_1(t)_{33} = 1, \tag{21}$$

$$K_1(t)_{44} = b_{44}^1 + b_{44}^2 \cos\left(\sqrt{|\Omega_1|^2 + |\Omega_2|^2 + |g|^2} \cdot t\right), \tag{22}$$

$$K_1(t)_{55} = b_{55}^1 + b_{55}^2 \cos(\alpha_1 \cdot t) + b_{55}^3 \cos(\alpha_2 \cdot t), \tag{23}$$

$$\alpha_{1,2} = \sqrt{\frac{1}{2}\left[|\Omega_1|^2 + |\Omega_2|^2 + 2|g|^2 \pm \sqrt{\left(|\Omega_1|^2 + |\Omega_2|^2\right)^2 + 4|g|^2|\Omega_2|^2}\right]},$$

$$K_1(t)_{66} = b_{66}^1 + b_{66}^2 \cos(\beta_1 \cdot t) + b_{66}^3 \cos(\beta_2 \cdot t), \qquad (24)$$

$$\beta_{1,2} = \sqrt{\frac{1}{2}\left[|\Omega_1|^2 + |\Omega_2|^2 + 2|g|^2 \pm \sqrt{\left(|\Omega_1|^2 + |\Omega_2|^2\right)^2 + 4|g|^2|\Omega_1|^2}\right]},$$

$$K_1(t)_{77} = b_{77}^1 \cos(\gamma_1 \cdot t) + b_{77}^2 \cos(\gamma_2 \cdot t) + b_{77}^3 \cos(\gamma_3 \cdot t), \qquad (25)$$

$$\gamma_1 = |g|, \; \gamma_{2,3} = \sqrt{\frac{1}{2}\left[|\Omega_1|^2 + |\Omega_2|^2 + 2|g|^2 \pm \sqrt{\left(|\Omega_1|^2 + |\Omega_2|^2\right)\left(|\Omega_1|^2 + |\Omega_2|^2 + 4|g|^2\right)}\right]}.$$

Here $b_{ss}^l$ is the corresponding element of matrix $B_l$ obtained from the system (18). The analytical expressions for other operators $K_{k\neq 1}(t)$ may be found as linear combinations of $K_1(t)$ and its derivatives. For instance and for our further purpose also write out the operator of quantum jump $K_2(t)$ which may be expressed as follows:

$$K_2(t) = \frac{ia_a}{g_a^*} \frac{\dot{K}_1(t)}{a_a a_a^+}. \qquad (26)$$

### 4. CPS gate implementation

In this section we introduce the results followed from analyze done in section 3. It was mentioned above that $Q_\pi^{(3)}$ may be implemented with operator $K_1(t)$ in use when following conditions on its matrix elements $K_1(t)_{ij}$ at time $t_{int}$ are satisfied:

$$K_1(t_{int})_{11} = K_1(t_{int})_{22} = \ldots = K_1(t_{int})_{77} = 1, \; K_1(t_{int})_{88} = -1, \; K_1(t_{int})_{i,j\neq i} = 0. \qquad (27)$$

Here we will find the values of controlled parameters and time interaction which being inserted in (22) - (25) give required operator. Also the fidelity and probability of CPS transformation will be estimated. To obey (27) there are Rabi frequencies, coupling constants and time interaction in our disposal. It should be noted that a part of these conditions satisfied automatically (21). For simplicity we assume the following relation between parameters of quantum and classical fields:

$$|\Omega_1| = |\Omega_2| = |\Omega| \gg |g_a| = |g_b| = |g_c| = |g|. \qquad (28)$$

This gives the approximate analytical expressions for elements of $K_1(t)$:

$$K_1(t)_{55} = 1 + \frac{|g|^2}{2|\Omega|^2}\left[\cos(\sqrt{2}|\Omega|t) - 2\right], \qquad (29)$$

$$K_1(t)_{66} = K_1(t)_{77} = 1 + \frac{|g|^2}{2|\Omega|^2}\left[\cos(\sqrt{2}|\Omega|t) + \cos\left(\frac{|g|}{\sqrt{2}}t\right) - 2\right], \qquad (30)$$

$$K_1(t)_{88} = \cos\left(\frac{|g|^2}{\sqrt{2}|\Omega|}t\right) + \frac{|g|^2}{2|\Omega|^2}\left[\cos(\sqrt{2}|\Omega|t) - \cos\left(\frac{|g|^2}{\sqrt{2}|\Omega|}t\right)\right]. \qquad (31)$$

Using these results we may estimate time interaction $t_{int}$ between atom and field in the cavity which required for operation $Q_\pi^{(3)}$. Satisfying Eq. (27) in zero-order of $|g|/|\Omega|$ one can obtain:

$$t_{int} = \frac{\sqrt{2}\pi|\Omega|}{|g|^2}(1+2k),\; k=0,1,2,\ldots \qquad (32)$$

Let us consider the probability of this operation and its fidelity after interaction lasted the time $t_{int}$. The probability of realization $Q_\pi^{(3)}$ equals to the probability to detect atom in its initial state $|1\rangle_A$. It may be written as follows:

$$P(t) = Tr_F\{K_1(t)\rho_f(0)K_1^+(t)\} = \sum_{s=0}^{7} \rho_f(0)_{ss}|K_1(t)_{ss}|^2. \qquad (33)$$

Here $\rho_f(0)$ is the arbitrary density matrix of cavity field at initial time. Substituting (32) into (33) and using analytical expressions for $K_1(t)$ we obtain that $P(t_{int})$ ($k=0$) differs from unity by the value:

$$1 - P(t_{int}) = \frac{2|g|^2}{|\Omega|^2}\sum_{s=4}^{7}\rho_f(0)_{ss}, \qquad (34)$$

where we have averaged the fast oscillations on frequencies $|\Omega|$. The difference $1-P(t)$ is depicted on Fig. 2 in the case then $\rho_f(0) = |\xi\rangle_F\langle\xi|$ corresponds to the initial field state $|\xi\rangle_F = |1\rangle_a(|00\rangle + |01\rangle + |10\rangle + |11\rangle)_{bc}$.

For conditional fidelity we can write:

$$F(t) = \sqrt{Tr_F(\rho_Q \cdot \rho_f^c(t))}. \qquad (35)$$

Here $\rho_f^c(t)$ is conditional density matrix of field after measurement:

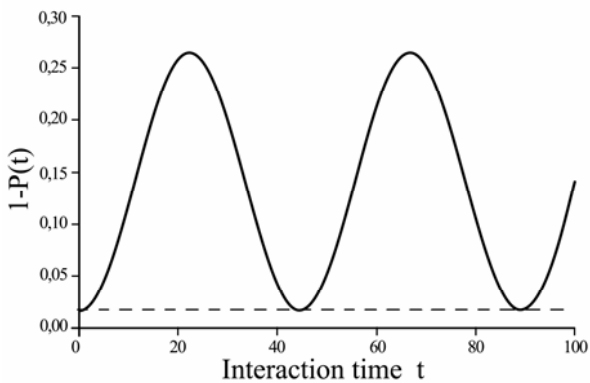 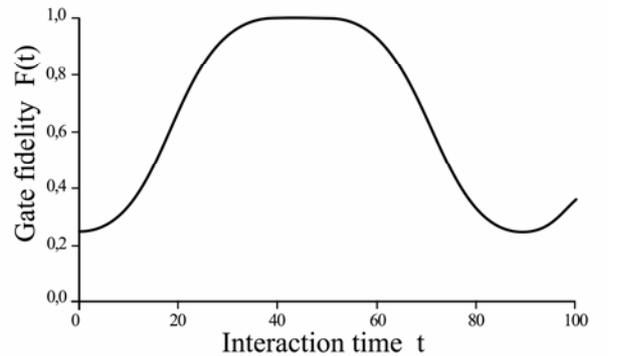

**Fig. 2:** Probability $1-P$ not to detect atom in ground state $|1\rangle_A$ as function of atom-cavity interaction time $|\Omega|=10|g|$, $\Delta_j=0$ (solid line) and its analytical estimation for $t=t_{int}$ obtained from Eq. (34) (dashed line).

**Fig. 3:** Fidelity of quantum operation $Q_\pi^{(3)}$ as a function of time $|\Omega|=10|g|$, $\Delta_j=0$.

$$\rho_f^c(t) = \frac{K_1(t)\rho_f(0)K_1^+(t)}{Tr_F\{K_1(t)\rho_f(0)K_1^+(t)\}}, \qquad (36)$$

and $\rho_Q = Q_\pi^{(3)}\rho_f(0)Q_\pi^{(3)+}$ is a result of ideal $Q_\pi^{(3)}$ transformation of initial state $\rho_f(0)$. It is simple to verify that $F(t_{int}) = 1$ up to second order. Fig. 3 shows the time dependent fidelity obtained from numerical counting with $\rho_f(0) = |\xi\rangle_F \langle\xi|$.

## 5. EPR pair generation

Here we represent some extra facility of the system described above. Namely, we are going to investigate the conditional field state which appears after atom is detected in state $|2\rangle_A$. It corresponds to conditional field evolution driving by Kraus operator $K_2(t)$ describing quantum jump. We will show how the entangled EPR state may be generated from the following initially disentangled state:

$$|\varphi\rangle_F = \frac{i}{2}(|100\rangle_F + |101\rangle_F + |110\rangle_F + |111\rangle_F) = |1\rangle_a \otimes \frac{i}{\sqrt{2}}(|0\rangle_b + |1\rangle_b) \otimes \frac{1}{\sqrt{2}}(|0\rangle_c + |1\rangle_c). \qquad (37)$$

Here we use indices a, b and c for corresponding mode of cavity field. This state may be obtained after switching off the classical modes and using the methods described in [19]. To generate EPR state we consider the time dependent elements of operator $K_2(t)$ in Fock basis under the following assumption:

$$|\Omega_1| = |\Omega_2| = |g_a| = |g_b| = |g_c| = |g|. \qquad (38)$$

Then the nonzero elements of $K_2(t)$ may be written as follows:

$$K_2(t)_{15} = -\frac{i}{\sqrt{3}}\sin(\sqrt{3}|g|t), \qquad (39)$$

$$K_2(t)_{26} = K_2(t)_{37} = -\frac{i}{4}\left\{\sqrt{2+\sqrt{2}}\sin\left(\sqrt{2+\sqrt{2}}|g|t\right) + \sqrt{2-\sqrt{2}}\sin\left(\sqrt{2-\sqrt{2}}|g|t\right)\right\}, \qquad (40)$$

$$K_2(t)_{48} = -\frac{i\sqrt{2}}{\sqrt{3}}\sin\left(\frac{\sqrt{3}|g|t}{\sqrt{2}}\right)\cos\left(\frac{|g|t}{\sqrt{2}}\right). \qquad (41)$$

To get the EPR state from initial state $|\varphi\rangle_F$ the following conditions need to be satisfied:

$$K_2(t)_{15} = K_2(t)_{48} = 0, \qquad (42)$$

$$K_2(t)_{26} = K_2(t)_{37} \neq 0. \qquad (43)$$

Then the conditional field state gives required entanglement between second and third modes:

$$|EPR\rangle_F = \frac{|010\rangle_F + |001\rangle_F}{\sqrt{2}} = |0\rangle_a \frac{|10\rangle_{bc} + |01\rangle_{bc}}{\sqrt{2}}. \qquad (44)$$

Eq. (42) and (43) give the restriction for time interaction duration. The intersection of these sets is empty, but it is possible to choose the time, which satisfies (42) and (43) approximately. For instance we take:

$$t_{int} = \frac{6\pi}{\sqrt{3}|g|}. \tag{45}$$

From here we can obtain the probability to detect atom in state $|2\rangle_A$ and fidelity of corresponding field transformation after interaction time $t_{int}$: $P \approx 0.2$, $F \approx 0.99$.

## 6. Discussion and Conclusion

Here we considered the implementation of probabilistic conditional phase shift operation $Q_\pi^{(3)}$ in high-Q cavity of QED configuration. It transforms three-qubit state, which encoded into the photon numbers of three quantum cavity modes. To perform this operation the conditional field state evolution determined by result of atomic state detection were used. We choose the Kraus operators formalism to describe this evolution. It helps to simplify the analysis of field evolution due to absence of necessity considering every invariant subspace of atom-field states separately (see [20, 22]). Also it allows exact analytical expressions for some Kraus operators in resonant approach. Due to the fact that photon storage time (1 ms) in cavity is exceed sufficiently atom cavity interaction duration (~10 μs) the relaxation channels were ignored.

The set of Kraus operators corresponding to different atomic measurement results consists of conditional evolution operator $K_1(t)$ and operators which describe different quantum jumps $K_{j \neq 1}(t)$. It was shown that $Q_\pi^{(3)}$ transformation may be implemented using the operator $K_1(t)$ which describes field evolution then atom being detected in its ground state $|1\rangle_A$. This operator is diagonal in the basis of Fock states. Consequently it is very convenient to realize the phase operations representing the qubits as photon number of cavity mode. But in this representation the realization of Hadamard operation is nontrivial problem because of the cavity mode energy in this case doesn't conserve. This is an obstacle on the way of CCNOT gate implementation. Technique which used here allows the Hadamard operation with only 50% probability. Though this problem was solved successfully for single cavity in [20] where Hadamard transformation was obtained in three steps. At the first and third steps atom interacts with quantum cavity field, while at the second step quantum mode are switched off and atom interacts with mode of classical field. Another way of realizing the Hadamard operation based on several cavities and "atomic mirror" is suggested in [21].

## Acknowledgments

The work is supported by government contracts P689_NK-526P and R&D PK10186. Project 2.1.1/4215, by the Grant of the St. Petersburg Government for Students and Postgraduates of Higher Education Schools and Academic Institutes of St. Petersburg.